\documentclass[sigconf,nonacm]{acmart}

\AtBeginDocument{%
\providecommand{\BibTeX}{{%
\normalfont B\kern-0.5em{\scshape i\kern-0.25em b}\kern-0.8em\TeX}}}

\setcopyright{acmcopyright}

\copyrightyear{2023}
\acmYear{2023}
\setcopyright{acmlicensed}\acmConference[IoTDI '23]{International Conference
on Internet-of-Things Design and Implementation}{May 9--12, 2023}{San
Antonio, TX, USA}
\acmBooktitle{International Conference on Internet-of-Things Design and
Implementation (IoTDI '23), May 9--12, 2023, San Antonio, TX, USA}
\acmPrice{15.00}
\acmDOI{10.1145/3576842.3582365}
\acmISBN{979-8-4007-0037-8/23/05}

\acmConference[IoTDI'23]{ACM/IEEE Conference on Internet of Things Design and Implementation}{May 09--12, 2023}{San Antonio, Texas}
\usepackage{subcaption}
\usepackage{multirow}

\usepackage{eso-pic}
\usepackage{url}
\AddToShipoutPictureBG*{
  \AtPageUpperLeft{%
    \put(0,-40){\raisebox{15pt}{\makebox[\paperwidth]{\begin{minipage}{21cm}\centering
      \textcolor{gray}{This article has been accepted for publication in the proceedings of the 8th ACM/IEEE Conference on Internet of Things Design and Implementation \\Content may change prior to final publication. DOI: 10.1145/3576842.3582365} 
    \end{minipage}}}}%
  }
  \AtPageLowerLeft{%
    \raisebox{25pt}{\makebox[\paperwidth]{\begin{minipage}{21cm}\centering
      \textcolor{gray}{ \copyright 2023 ACM.  Personal use of this material is permitted.  Permission from ACM must be obtained for all other uses, in any current or future media, including reprinting/republishing this material for advertising or promotional purposes, creating new collective works, for resale or redistribution to servers or lists, or reuse of any copyrighted component of this work in other works.
      }
    \end{minipage}}}%
  }
}

\begin{document}

    \title[In-Ear-Voice: Towards Milli-Watt Audio Enhancement With Bone-Conduction Microphones]{In-Ear-Voice: Towards Milli-Watt Audio Enhancement With Bone-Conduction Microphones for In-Ear Sensing Platforms}


	\author{Philipp Schilk}
	\orcid{0000-0002-0487-9513}
	\authornote{Both authors contributed equally to this research.} \affiliation{%
	\institution{Eidgenössische Technische Hochschule Zürich (ETHZ)} \city{Zürich} \country{Switzerland}}
    \email{schilkp@ethz.ch}

	\author{Niccolò Polvani}
	\orcid{0000-0003-0720-050X}
	\authornotemark[1]
    \affiliation{\institution{École polytechnique fédérale \\ de Lausanne (EPFL)} \city{Lausanne} \country{Switzerland}}
    \affiliation{\institution{Logitech Europe} \city{Lausanne} \country{Switzerland} } 
    \email{npolvani@logitech.com}

	\author{Andrea Ronco}
	\orcid{0000-0003-1672-8672}
	\affiliation{\institution{Eidgenössische Technische Hochschule Zürich (ETHZ)} \city{Zürich} \country{Switzerland}}
    \email{andrea.ronco@pbl.ee.ethz.ch}

	\author{Milos Cernak}
	\orcid{0000-0002-5569-9491}
	\affiliation{\institution{Logitech Europe} \city{Lausanne} \country{Switzerland}}
    \email{mcernak@logitech.com}

	\author{Michele Magno}
	\orcid{0000-0003-0368-8923}
	\affiliation{\institution{Eidgenössische Technische Hochschule Zürich (ETHZ)} \city{Zürich} \country{Switzerland}}
    \email{michele.magno@pbl.ee.ethz.ch}

	\renewcommand{\shortauthors}{Schilk and Polvani, et al.}

	\begin{abstract}
The recent ubiquitous adoption of remote conferencing has been accompanied by omnipresent
frustration with distorted or otherwise unclear voice communication.
Audio enhancement can compensate for low-quality input signals from, for example, small true wireless earbuds,
by applying noise suppression techniques.
Such processing relies on voice activity detection (VAD)
with low latency and the added capability of discriminating the wearer's
voice from others - a task of significant computational complexity. The tight energy budget of devices as small as modern
earphones, however, requires any system attempting to tackle this problem to do so with
minimal power and processing overhead, while not relying on speaker-specific voice samples 
and training due to usability concerns.

This paper presents the design and implementation of a custom research platform 
for low-power wireless earbuds based on novel, commercial, MEMS bone-conduction microphones. 
Such microphones can record the wearer's speech with much greater
isolation, enabling personalized voice activity
detection and further audio enhancement applications. Furthermore, the paper accurately evaluates a proposed low-power personalized speech detection algorithm based on bone conduction 
data and a recurrent neural network running on the implemented research platform. This algorithm 
is compared to an approach based on traditional microphone input.
The performance of the bone conduction system, achieving detection of speech within 12.8ms
at an accuracy of 95\% is evaluated. Different SoC choices are contrasted, 
with the final implementation based on the cutting-edge
Ambiq Apollo 4 Blue SoC achieving 2.64mW average power consumption at 14uJ per inference, 
reaching 43h of battery life on a miniature 32mAh li-ion cell and without duty cycling.
\end{abstract}

	\maketitle

	\section{Introduction}

In the last two decades, a significant transformation in remote vocal interaction has occurred.
The world is getting more connected and globalized, creating
more opportunities for people at a distance to interact. In addition, the diffusion
of consumer electronics and increase in bandwidth availability enables more frequent
calls and video conferences, with little limitations on physical location \cite{increase_in_remoteconf}. The COVID-19 pandemic further accelerated the natural adoption of such technologies,
making them a fundamental organizational resource in many businesses today \cite{conference_after_covid}. For this
reason, a lot of resources are being spent by both industry and academia 
attempting to increase the quality of the user's experience during remote
conferencing. However, audio quality is still a common challenge, with distorted audio making 
communication ineffective, frustrating, or even impossible \cite{voice_frustration}.

Simultaneously, small true-wireless earbuds have become more and more ubiquitous, further exacerbating the problem:
Their small microphone is placed indirectly and at a substantial distance from the wearer's mouth, reducing the audio quality and 
increasing the amount of extraneous noise captured, including both other nearby speakers and common 
environmental noise.
While audio enhancement and filtering is a well-studied field, this new generation of miniaturized
devices entail extremely challenging boundary conditions for any attempt at audio improvement \cite{hearables_dsp}.
Such "Hearable" devices are usually based on rather limited hardware, providing low computational 
resources in exchange for low power consumption in the range of a few milli-watts - 
a necessity, given the tiny amount of battery capacity that can be feasibly packaged in such small 
devices \cite{hearables_overview}.

Data transmission is also very pricey in these systems \cite{wireless_costly}, with wireless interfaces
characterized by the usual trade-off between power consumption and bandwidth. This means that offloading data is very expensive, and puts the
generous computational resources of the host devices, be that a laptop or a smartphone, at an unreachable distance.
Furthermore, such transmission would introduce additional delay, worsening the user 
experience.

For this reason, data transmission is usually avoided unless strictly necessary,
and the processing is kept local, requiring the processing algorithms
to be implemented on constraint devices with limited storage and run-time memory,
basic processing capabilities, and low clock speeds \cite{limited_embedded_capabilites}.
These challenges are even more critical when machine learning techniques are employed as a part of the signal 
processing pipeline. With traditional models designed to
run on powerful hardware spanning from personal computers to data centers, many 
state-of-the-art processing methods are simply viable in this domain \cite{lp_noisered}.

Tiny Machine Learning (TinyML) attempts to address the issue of deploying machine learning
solutions on embedded hardware. TinyML models are developed with
a different mindset, considering the hardware limitations from the
beginning and reflecting these constraints in the network architecture \cite{tinyml_survey}. A lot
of effort is also put into appropriate pre-processing and feature extraction,
which can significantly benefit the performance of the models with a lower
computational cost. In addition, different techniques are used to improve the
hardware utilization and efficiency, by the means of Quantization \cite{tinyml_quant}, Network
Pruning \cite{tinyml_prune}, Transfer-Learning \cite{tinyml_transfer}, and Compression \cite{tinyml_compr}. 
This enables an application-specific model small enough for hearable platforms to be developed.

The recent emergence of low-power MEMS bone conduction microphones \cite{ac_bc_fusion} provides an exciting 
opportunity for novel enhancement approaches. Their inherent improved isolation from external noise and different recording 
characteristics provide another representation of the same signal.
In combination with TinyML techniques, this has the potential to greatly reduce the processing required
by providing information that previously had to be learned from much less expressive features at great computational cost.

The main contributions of this paper are:
\begin{itemize}
	\item Design and implementation of a hardware platform for in-ear sensing 
          research featuring novel sensing and cutting-edge signal processing capabilities.
    \item Development and evaluation of an ultra-low-power, bone-conduction based, 
          Personalized Voice Activity Detection (pVAD) module with an exploration of
          further possible power savings.
	\item Comparison of the industry-standard Nordic NRF5340 to the cutting edge 
          Ambiq Apollo 4 Blue for ultra-low-power edge processing.
\end{itemize}

	\section{Related Work}
Reliable Voice Activity Detection (VAD) is a core step toward speech enhancement.
It provides an essential data point for further processing and can be employed 
as a gating mechanism for more intensive processing, saving precious computing and 
power resources.

\subsection{Low power VAD}

Ultra-low power VAD is a well-studied application and has been attempted with 
an extensive array of techniques.
FPGA-based solutions have been demonstrated with 560$\mu$W of power consumption 
\cite{ulp_vad_fpga} \cite{ulp_vad_fpga_2}, while neuromorphic implementations have 
achieved power consumption as low as 30$\mu$W \cite{ulp_vad_logi1} and 4$\mu$W \cite{ulp_vad_logi2}, 
when disregarding the power-intensive support hardware required to run such inferences. 
A Deep Neural Network approach supported by custom silicon has been shown to require only 22.3$\mu$W \cite{ulp_vad_silicon}.
State-of-the-art implementations relying on analog pre-processing and feature extraction 
followed by a deep neural network are reaching power consumptions as low as 1$\mu$W and 142nW \cite{ulp_vad_analog1} \cite{ulp_vad_analog2}.

The applicability of such general-purpose VAD to speech enhancements for hearables is, however, questionable. 
Voice communication is not only plagued by general environmental noise but also by the nearby speech from other people.
VAD that detects speech but cannot discriminate a target speaker from others, such as the 
examples above, would be ineffective in applications attempting to enhance only the voice 
of the wearer of a given earbud.

\subsection{Personalized VAD (pVAD)}
Personalized voice activity detection (pVAD) has also been the subject of in-depth research.
Attempts commonly utilize a generic VAD module followed by speaker diarization 
in the form of a model trained on sample target speech  or 
train a VAD module on samples of target speech directly \cite{pvad_with_training} \cite{pvad_with_training_2} \cite{pvad_two_steps}.
Such approaches tend to be very complex, some requiring networks with up to 130K parameters. Furthermore,
they require sufficient speech samples from the target voice (enrollment data).
Other novel approaches propose methods that enable training without such enrollment data. This is achieved by developing a model which,
given a sample of target speech and noisy audio, can detect if the target speaker is present in the noisy audio  \cite{pvad_with_training_augment}. 
Requiring only a small sample of target 
speech is much more practical but still requires active interaction from the user. 
While effective, these models are necessarily very complex, making them impractical for edge computing. 
This could be overcome via the offloading of computation to a more capable device, but the power required 
to transmit the raw audio data and the latency introduced makes this impractical.

\subsection{Bone Conduction Microphones}
MEMS-based bone conduction microphones provide a possible avenue for innovation. By relying on a direct coupling to the wearer's head, they are capable of
capturing speech with a much higher level of isolation at the cost of somewhat unnatural sounding recordings \cite{bc_drawbacks}. 
Using this unique property, they have already been employed in applications such as audio enhancement by fusion with a standard air conduction signal \cite{ac_bc_fusion}, 
human sound classification \cite{bc_classify}, and pitch detection \cite{bc_pitch}. While they require direct physical contact to be effective, the ear has been 
shown to be an effective site for bone conduction microphone placement \cite{bc_loc_comparison}, making them a natural fit for integration in tiny earbuds. Furthermore, 
the different frequency response imparted on the bone-conducted signal, while somewhat detrimental to applications targeting speech enhancement and reconstruction, provides 
a valuable differentiation between the wearer's voice and adjacent speakers, enabling personalized voice activity 
detection without enrollment data and significantly less processing overhead.

This paper presents the design of an energy-efficient and low-power earbud that exploits novel bone conduction sensors.
Moreover, we design and implement a tiny machine learning algorithm for speech detection that can run on a state-of-art
ARM Cortex-M4F core embedded in a Bluetooth low energy system on chip (SoC). The paper accurately presents the quality of
the implementation and energy efficiency. To the best of our knowledge, this is the first paper that presents the design and
implementation of such a device combining the bone conduction microphone and tiny machine learning to have a truly wireless and
intelligent wearable in-ear device.

	\section{System Design}
\label{sec:hw}

\begin{figure}[b]
	\centering
	\includegraphics[scale=0.70]{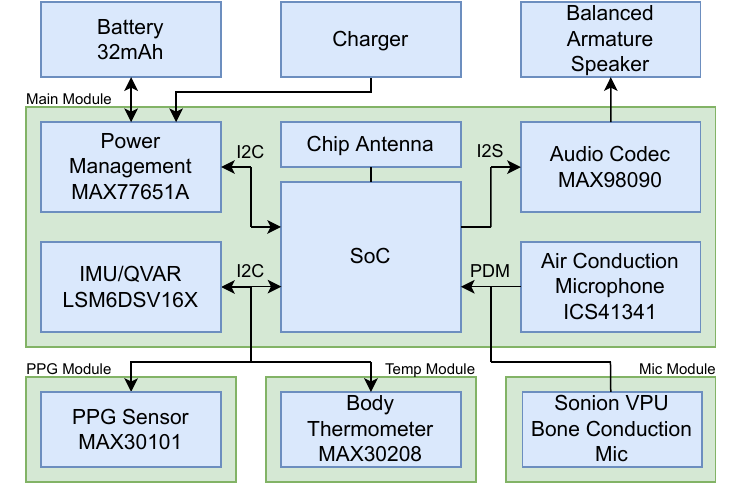}
	\caption{System architecture.}
	\label{fig:sys_arch}
\end{figure}

With no off-the-shelf development platform available for in-ear sensing research, 
a custom earbud was developed. It is a modular design, with a central module housing 
power infrastructure, the main SoC, and some sensors, with additional sensors placed 
on small external modules which can be freely installed inside an earphone case. This enables
flexibility during development, as the exact type and location of sensors can easily be adjusted.
A system overview is shown in figure \ref{fig:sys_arch}. All components are integrated on 
custom, miniaturized PCBs. Uniquely, this platform provides a modular and flexible platform 
for fast in-ear sensing research at a level of integration usually reserved for 
commercial products.

\begin{figure*}[t]
    \begin{minipage}[b]{0.32\textwidth}
        \includegraphics[width=\textwidth]{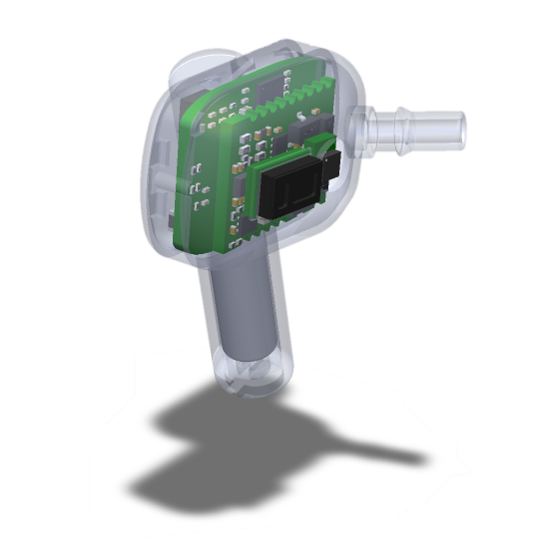}\\
        \subcaption{}
        \label{fig:des1}
    \end{minipage}%
    \begin{minipage}[b]{0.32\textwidth}
        \includegraphics[width=\textwidth]{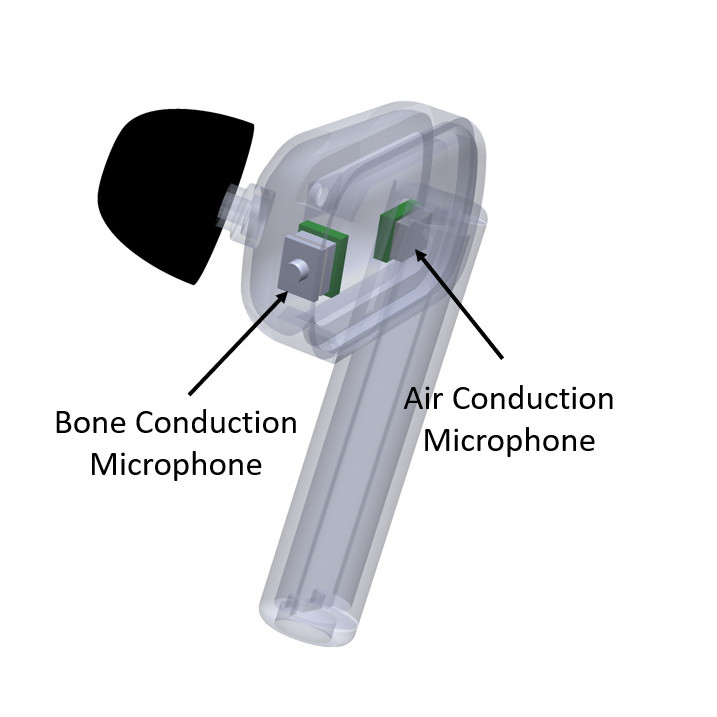}\\
        \subcaption{}
        \label{fig:des2}
    \end{minipage}%
    \begin{minipage}[b]{0.32\textwidth}
        \includegraphics[width=\textwidth]{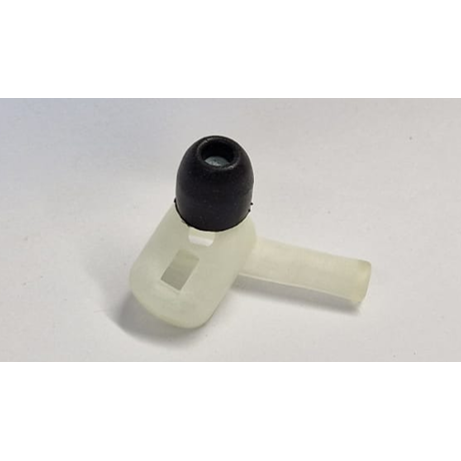}\\
        \subcaption{}
        \label{fig:des3}
    \end{minipage}%
    \caption{Design renders showing internal circuitry (\ref{fig:des1}), microphone placement (\ref{fig:des2}), and the 3D-printed case (\ref{fig:des3}).}
    \label{fig:design}
\end{figure*}

\begin{figure*}[t]
    \begin{minipage}[b]{0.32\textwidth}
        \centering
        \includegraphics[width=0.8\textwidth]{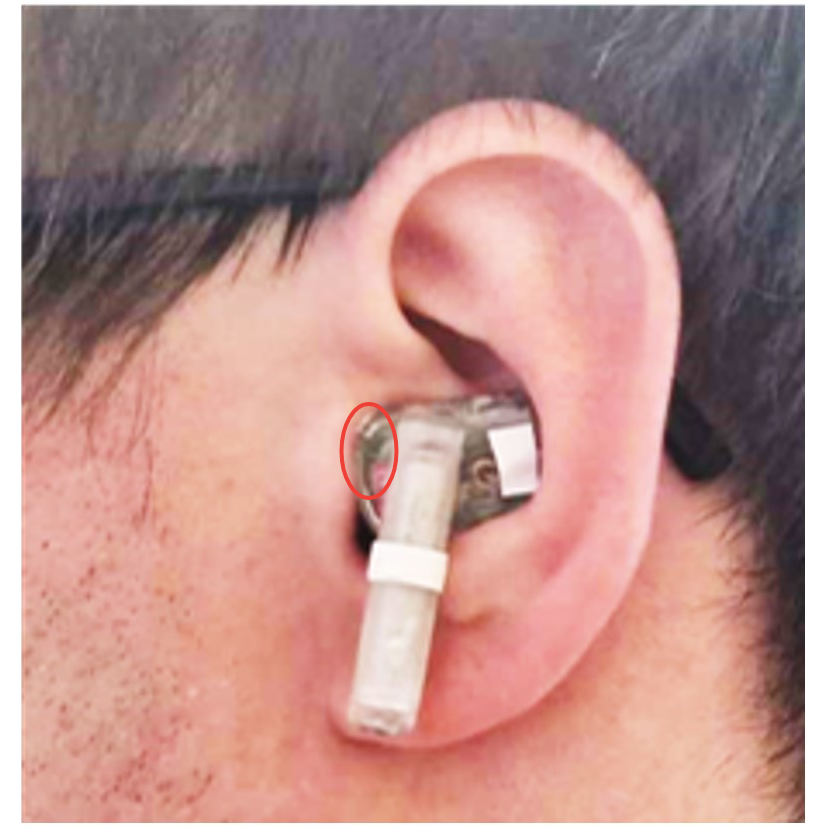}\\
        \subcaption{}
        \label{fig:dev1}
    \end{minipage}%
    \begin{minipage}[b]{0.32\textwidth}
        \centering
        \includegraphics[width=0.8\textwidth]{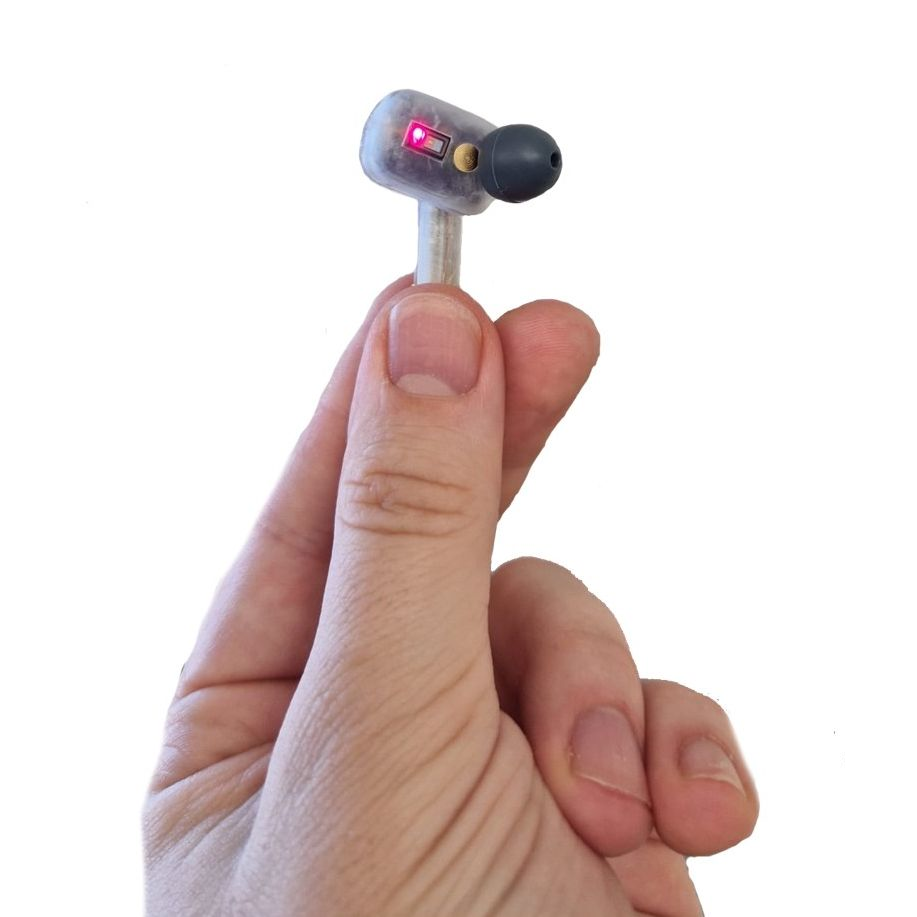}\\
        \subcaption{}
        \label{fig:dev2}
    \end{minipage}%
    \begin{minipage}[b]{0.32\textwidth}
        \centering
        \includegraphics[width=0.8\textwidth]{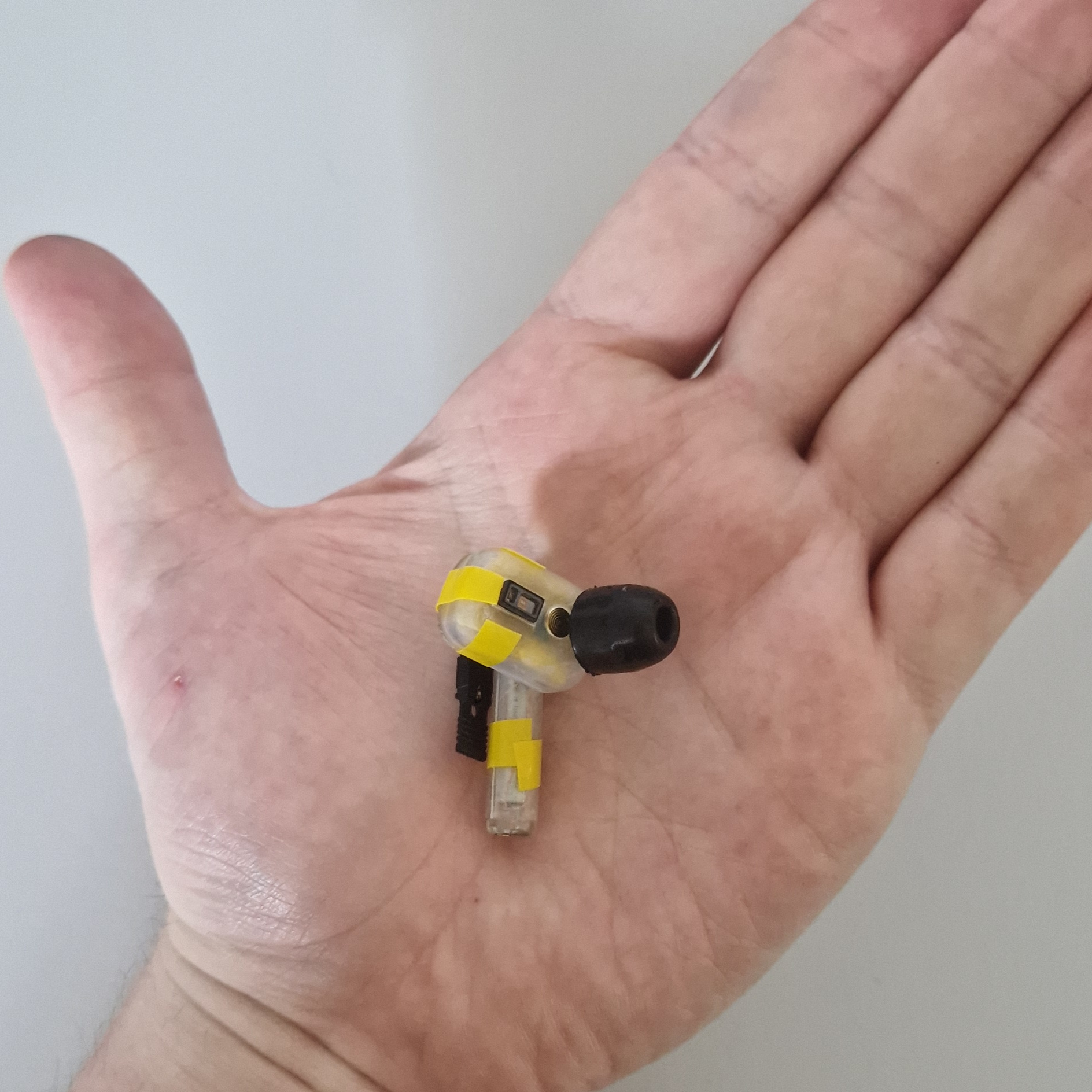}\\
        \subcaption{}
        \label{fig:dev3}
    \end{minipage}%
    \caption{The finished device. The position of the bone conduction microphone is indicated in figure \ref{fig:dev1}.}
    \label{fig:device}
\end{figure*}

\paragraph{Case:}
The earbud is housed in a plastic case, shown in figures \ref{fig:des1} - \ref{fig:des3}. 
Designed to accommodate all sensors and electronics while being comfortable, the main corpus
measures 13 mm $\times$ 9.5 mmm $\times$ 21 mm. The stem housing the battery is 28 mm long.
The case was designed and manufactured in-house using a resin-based 3D printing process.

\paragraph{SoCs:}
Two main modules were developed, identical in all but the central SoC. The first is based around the
Nordic NRF5340, a dual-core Cortex-M33 processor featuring an application core with a maximum clock speed of 128 MHz, 1 MB Flash, 512 kB RAM, and 
an integrated BLE5.3 transceiver managed by a dedicated network processor. It is one of the few SoCs models available with software support for 
BLE LE Audio, and reasonably representative of processors found in comparable consumer devices.

While the NRF5340 is very power efficient, developments in sub-threshold CMOS techniques have seen a new class of 
devices with best-in-class power consumption specifications become available \cite{ambiq_subth}. One such device is 
the cutting-edge Ambiq Apollo 4, kindly provided by Ambiq before their public release. Previous 
generations of the Apollo have already proven themselves to be excellent in power-efficient neural network inference \cite{marco_survey},
but no concrete evaluation of the new device exists. For this reason, a second main module was developed based on this SoC.

The Apollo 4 Blue is a dual-core SoC with a Cortex-M4 application core running at a clock speed of up to 196 MHz, 2 MB non-volatile storage, and  
over 2 MB RAM. It features a BLE5.1 transceiver with a dedicated Cortex-M0 network core and is specified at an impressive 5 $\mu A$/MHz.

\paragraph{Microphones:} The earbud features both an air-conduction and bone-conduction microphone.
The air conduction microphone is an ICS41341 low-power MEMS microphone, facing outwards on the back of the device.
The bone conduction microphone is a VPU14DB01 MEMS-based VPU microphone, kindly provided by Sonion. By the 
manufacturer's recommendations, it was placed facing forward towards the wearer's tragus.
The exact location of both microphones is shown in figure \ref{fig:des2}, and the position of the 
bone conduction microphone again highlighted in figure \ref{fig:des1}.

\paragraph{Other Sensors:}
Earbuds are in automatic intimate contact with the wearer. Therefore, they provide a unique opportunity 
for continuous, ubiquitous vital sign monitoring without wearing additional, possibly uncomfortable,
sensors. The earbud developed here also serves as an in-ear vital sign monitoring research platform.
For this reason, a collection of other health sensors are also available, including body temperature,
optical blood oxygen concentration, and heart rate sensing. These are the focus of a separate publication \cite{vitalpod}.

\paragraph{Battery and Power Management:}
The earbud is powered by a modern Panasonic CG-425A 32mAh pin-type Li-Ion battery housed in the stem of the case, chosen 
for its high energy density and convenient form factor. Battery charging, monitoring, and voltage regulation are handled 
by a highly efficient and integrated MAX77651 power management IC.

\paragraph{Firmware:}
Custom, Real-Time-Operating-System (RTOS) based firmware was developed for both SoCs, using 
FreeRTOS and Zephyr RTOS for the Apollo and NRF, respectively. This simplifies software development and 
enables the processor to be automatically put into a low-power sleep mode during periods of inactivity.

	\section{Personalised VAD Algorithm}

A recurrent neural network-based binary classifier operating in the frequency domain is 
used to detect the presence of target speech.

\subsection{Input}

The bone conduction microphone features a Pulse-Density Modulation (PDM) interface, which was
filtered and decimated using dedicated hardware peripherals inside the microcontroller.
This yields a 16bit, 16kHz uncompressed mono audio stream.

\subsection{Feature generation}

The proposed model operates in the frequency domain. To facilitate this,
the input stream was first separated into 20ms frames with 50\%
overlap, yielding frames of 320 samples in length.
An appropriate hamming window (20ms at 16kHz sample rate) was applied to each frame,
which was then zero-padded to 512 samples - a requirement for later processing steps to be 
optimized using hardware acceleration.

A Fast-Fourier-Transform (FFT) was applied, and the magnitude of the output spectrum warped to 32 
frequency bin features using a Mel-scale mapping. This is similar to the method proposed in \cite{net_shape} and illustrated in
figure \ref{fig:active_frames}. The reduction in input size significantly reduces model complexity, while the Mel-scale limits the amount of 
information lost during this transformation.

\begin{figure}[t]
	\centering
	\includegraphics[width=0.5\columnwidth]{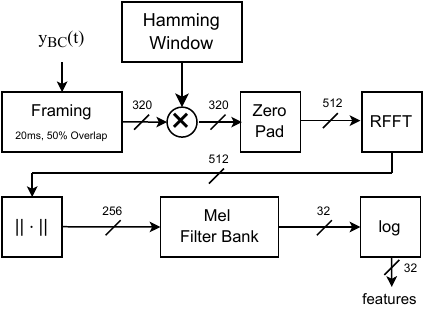}
	\caption{Scheme for Mel-scale feature extraction. RFFT refers to a hardware-optimized FFT implementation with purely real-valued inputs.}
	\label{fig:log_mel_features}
\end{figure}

\subsection{Model Architecture}
\label{sec:model_shape}

The model is a recurrent neural net with the following sequential architecture:

\begin{itemize}
	\item Two 1D Convolutional Layers in the frequency domain with kernel size 3,
		stride 2, and  respectively 16 and 32 output channels.

	\item Two Gated Recurrent Units with four activation units each.

	\item One fully connected hidden layer with 16 ReLU-activated output features.

	\item One fully connected layer with a single Sigmoid-activated output.
\end{itemize}

Such model architectures have been demonstrated to be very effective in similar applications
\cite{net_shape}. The two Gated Recurrent Units (GRUs) \cite{GRU} enable the
model to learn and recognize time-dependent information within the input signal.

The model has only approximately 5000 parameters, making it suitable for inference
even in performance and memory constraint environments. As is usual for binary classification, the output $y \in ]0,1[$ of the network can be converted 
to a binary label using a threshold.

	\section{Dataset Acquisition, Data Augmentation, and Model Training}
\subsection{Data Acquisition}
\label{sec:dataset}

Many datasets both for VAD and speaker identification \cite{dataset_2000_nist} \cite{dataset_realvad},
including some with common background noise samples \cite{dataset_qut_noise}, exist.
However, these feature speech recorded solely via air-conduction microphones.
Similarly, while some bone conduction datasets have been published \cite{dataset_bc_chinese},
no data featuring interference from adjacent speakers and other environmental
noise recorded by the bone conduction microphone is available.

This necessitated the creation of an in-house dataset, with the additional benefit of having all testing 
and training data collected using the same bone conduction
microphone model that would later be installed in the final device.

Three datasets were recorded, each featuring parallel recordings of in-ear bone
conduction (BC) microphones, in-ear air conduction (AC) microphones, and a reference external air conduction
microphone (AC-REF). Sonion VPU demo kits were used to record the two in-ear microphones, with the 
left earbud recording the air conduction and the
right earbud recording the bone conduction signal. A Rode M5 cardioid condenser
microphone served as the reference air conduction recording. All microphones
were recorded using a MOTU 4 audio interface.

In each dataset, the participant acts as the target speaker, whose voice activity
is to be detected, while simulated external speech and noise are to be rejected.

\begin{enumerate}
	\item \textbf{Target Speech Dataset}: Contains the target speaker's voice only.
		Speech from 20 participants (10 male, 10 female), reading text in 5
		languages (English 8, French 8, Italian 2, Spanish 1, Portuguese 1). In total
		approximately 2 hours of speech was collected. \label{ds_targspeech}

	\item \textbf{External Speech Dataset}: Contains external speech only. 10
		participants each listened to a total of 5 minutes of speech clips (\textasciitilde3 to 10
		seconds long) taken from the DNS challenge dataset \cite{dataset_DNS_challenge},
		without speaking. The speaker clips were played using two loudspeakers,
		placed approximately 1.5 m away from the participant. \label{ds_extspeech}

	\item \textbf{External Noise Dataset}: Contains external noise only. 10
		participants each listened to a total of 5 minutes of noise clips (\textasciitilde3 to 10
		seconds long) taken from the DNS challenge dataset \cite{dataset_DNS_challenge},
		without speaking. The noise clips were played using two loudspeakers, placed
		approximately 1.5 m away from the participant. \label{ds_extnoise}
\end{enumerate}

For dataset \ref{ds_targspeech}, the reference microphone was placed facing the
participant, while for dataset \ref{ds_extspeech} and \ref{ds_extnoise} the reference
microphone was placed to the right of the participant.

\subsection{Data Pre-Processing}
\label{sec:data_preproc}

\paragraph{Data Augmentation:}
To generate sufficient training samples, a procedure similar to \cite{dataset_qut_noise}
was employed, generating 30-second clips containing both target speech, external
speech and noise.

First, all recordings were re-sampled to 16 kHz to match the sample rate of the
microphones available in the final device. \\ Next, the final bone conduction
waveform that an earbud would capture, $y_{BC}(t)$, was modeled as the sum of
target speech $s_{BC}(t)$ and noise $\eta_{BC}(t)$, consisting of external speech,
$e_{BC}(t)$, and external noise, $\tilde{\eta} _{BC}(t)$:

\begin{gather}
	\label{eq:noisy_bc}%
	y_{BC}(t) = s_{BC}(t) + \gamma\cdot\eta_{BC}(t)\\%
	\eta_{BC}(t) = e_{BC}(t) + \tilde{\eta} _{BC}(t)
\end{gather}
with scaling factor $\gamma$ determining the relative signal and noise volumes.

For each clip, $s_{BC}$ was generated by randomly selecting a target speaker
from data set \ref{ds_targspeech}, and randomly cropping and concatenating without
separating intervals of continuous speech to form a 30-second sample. Clips from
dataset \ref{ds_extspeech} and \ref{ds_extnoise} were concatenated to form 30-second
samples of $e_{BC}$ and $\tilde{\eta} _{BC}$ respectively. These signals were then
combined (as per equation \ref{eq:noisy_bc}), selecting $\gamma$ to achieve a random target
$SNR \sim \mathcal{N}(1 5, 5)$:

\begin{equation}
	SNR_{BC, dB}= 10 \log_{10}\left(\frac{||s_{BC}||^{2}}{||\gamma\cdot\eta_{BC}||^{2}}\right)
\end{equation}

The final $y_{BC}$ was re-scaled to signal levels with $\mathcal{N}(-28, 10)$ dBFS
to train for level-independent detection. The training dataset created in this matter contains 160 hours
of simulated speech generated from 16 speakers and 160 hours of noise and external speech generated from 8
participants. The test dataset includes 5 hours of speech generated from the remaining four speakers, and 
five hours of noise and external speech generated using the recordings of 2 participants.

\begin{figure}[t]
	\centering
	\includegraphics[scale=0.60]{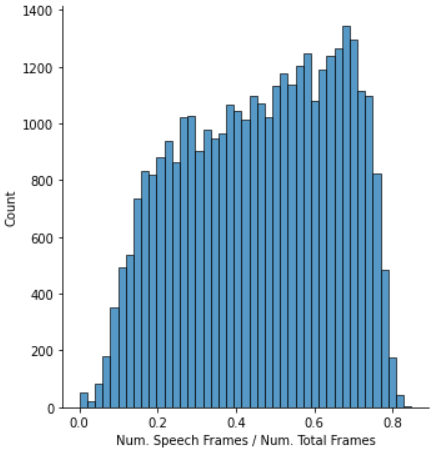}
	\caption{Distribution of active (speech) frames per total number of frames per file, obtained after processing the Speech Dataset.}
	\label{fig:active_frames}
	\vspace{-2em}
\end{figure}

Figure \ref{fig:active_frames} illustrates that the obtained dataset is evenly
divided between clips with low $(<25\%)$, medium $(25\% - 60\%)$ and high speech
content $(>60\%)$.

\paragraph{Label generation:}
The target voice activity labels $VAD(n)$ were generated from the reference air
conduction microphone recording $s_{AC,REF}$, corresponding to the $s_{BC}$ segments
in each clip:

\begin{gather}
	VAD(n) =%
	\begin{cases}
		1, & \mbox{if }||S_{AC,REF}(n)|| > T \\
		0, & \mbox{otherwise }
	\end{cases}
	\\%
	T = \min_{n}(||S_{AC,REF}(n)||) + \alpha \cdot \underset{n}{\mathrm{avg}}(||S_{AC,REF}
	(n )||)
\end{gather}

Where $\alpha$ was set to 0.3 by experimentation, and the vector $S_{AC,REF}(n)$ obtained by the STFT (Short-time Fourier Transform) of $s_{AC,REF}(t)$ with
20 ms frame size and $50\%$ overlap. A causal
averaging filter of 0.2 seconds was used to smooth the target labels.
It should be noted that the use of an air-conduction microphone with thresholding to generate
voice activity labels was only possible because all noise was recorded separately, and the air conduction 
microphone was facing the speaker. An ear-worn air conduction microphone, with its placement 
away from the speaker's mouth, records too much noise to reliably generate labels with such a
simplistic scheme.

\subsection{Training Procedure}
\label{sec:model_train} This being a binary classification task, Binary Cross Entropy
(BCE) was used as the loss function during training:
\begin{equation}
	\mathcal{L}_{BCE}(z) = - \frac{1}{N}\sum_{n}z_{n}\log(\hat{z}_{n}) + (1 - \hat{z}
	_{n}) \log (z_{n})
\end{equation}
where $z_{n}$ and $\hat{z}_{n}$ are the VAD labels and model's prediction
respectively, and $N$ is the number of input frames. An Adam optimizers
\cite{Adam} was used to update the model's weights, with an initial learning rate
of 1e-3. Each training epoch consisted of 2000 update steps, with batch size 8.
The learning rate was halved after three epochs without improvement on the test set,
and training was stopped after five epochs without improvement.

	\section{Results}

\begin{figure}[htbp]
	\centering
	\includegraphics[scale=0.50]{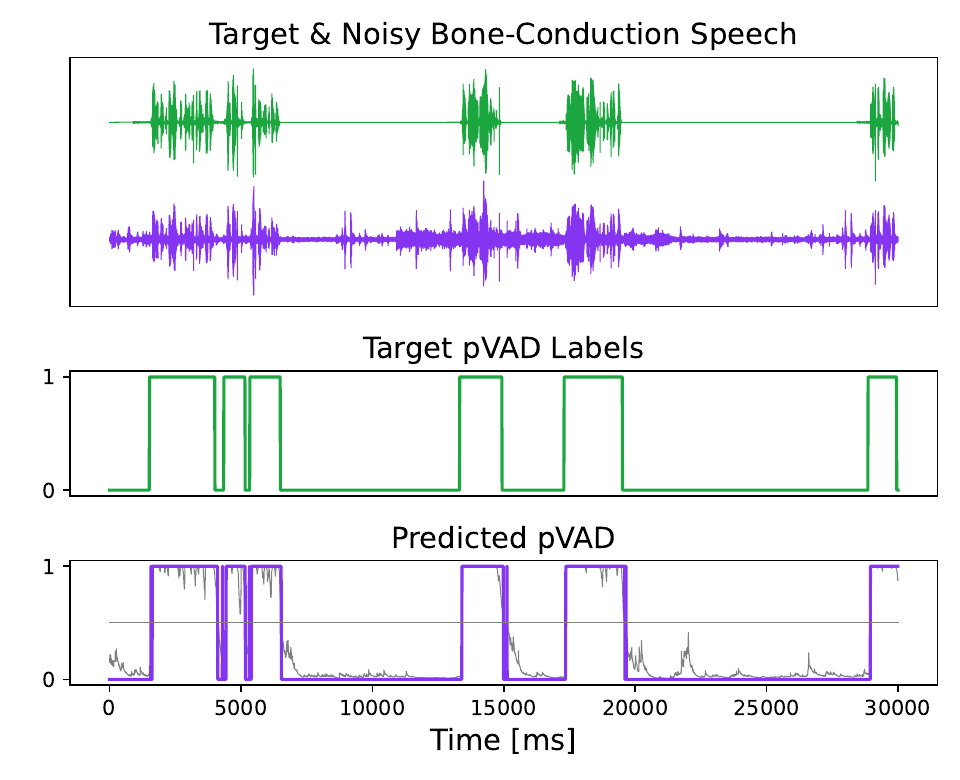}
	\caption{Visualization of an isolated target speech recording, noisy target speech recording, and model output.}
	\label{fig:nn_demo}
\end{figure}

%
\subsection{pVAD Evaluation}\label{sec:vad_eval}
To demonstrate not only the effectiveness of the proposed model (henceforth BC-pVAD)
but also the benefit a bone conduction microphone provides in this application, a
second model, AC-pVAD, was created. AC-pVAD employs precisely the same feature
extraction, network architecture, and training policy as BC-pVAD, but was instead
trained on noisy air conduction waveforms $y_{AC}$:
\begin{gather}
	\label{eq:noisy_air}y_{AC}(t) = s_{AC}(t) + \eta_{AC}(t)\\%
	\eta_{AC}(t) = e_{AC}(t) + \tilde{\eta}_{AC}(t)
\end{gather}
where $y_{AC}(t)$ is the noisy mixture, $s_{AC}(t)$ is the target user's speech,
$e_{AC}(t)$ is external speech and $\tilde{\eta}_{AC}(t)$ is external noise, all
recorded by the air-conduction microphone worn by the participants during the creation
of the datasets described in section \ref{sec:data_preproc}. Training data was
again generated using the procedure in section \ref{sec:data_preproc}, and the model
trained as described in section \ref{sec:model_train}. \\ \\ Both models will be
evaluated against the following metrics, as proposed in \cite{nist_1} \cite{nist_2}.

\begin{enumerate}
	\item Area under the receiver operating characteristic curve (AUC), which is
		threshold independent.

	\item Detection Cost Function (DCF), defined as:
		\begin{gather}
			DCF = 0.75 \cdot \textrm{Miss Rate}+ 0.25 \cdot \textrm{False Alarm Rate}\\
			\textrm{Miss Rate}= \dfrac{\textrm{total false negative time}}{\textrm{total speech time}}\\%
			\textrm{False Alarm Rate}= \dfrac{\textrm{total false positive time}}{\textrm{total non-speech time}}
		\end{gather}
		where speech time refers only to the target speech $s_{AC}$.

	\item Binary Accuracy (ACC) with a threshold of 0.5.
\end{enumerate}


\paragraph{Evaluation in Equal Environments:}
In general, a bone conduction microphone will be able to pick up target speech with
a much higher $SNR$ than what an air conduction microphone could achieve in the same
environment. The exact difference, however, is not constant: The fit of the earpiece,
and hence the coupling of target speech to the bone conduction microphone, varies
between different users, and even between different usage sessions by the same
user. 
To evaluate how both models would perform in an equivalent situation, the following
testing procedure was employed:

Given test clips of target speech and noise recorded simultaneously via the air conduction
and bone conduction microphone, a certain $SNR_{AC, dB}$ was fixed and the air
conduction speech and noise rescaled to this target SNR by selecting the coefficients $\alpha$
and $\beta$ accordingly:
\begin{gather}
	y_{AC}(t) = \alpha \cdot s_{AC}(t) + \beta \cdot \eta_{AC}(t) \\%
	SNR_{AC, dB}= 10 \log_{10}\left(
        \frac
        {||\alpha \cdot s_{AC}||^{2}}
        {||\beta \cdot \eta_{AC}||^{2}}
    \right)
\end{gather}

The same coefficients were then applied to the bone conduction data:

\begin{equation}
	y_{BC}(t) = \alpha \cdot s_{BC}(t) + \beta \cdot \eta_{BC}(t)
\end{equation}

\begin{figure*}[t]
    \begin{minipage}[b]{0.3\textwidth}
        \includegraphics[width=\textwidth]{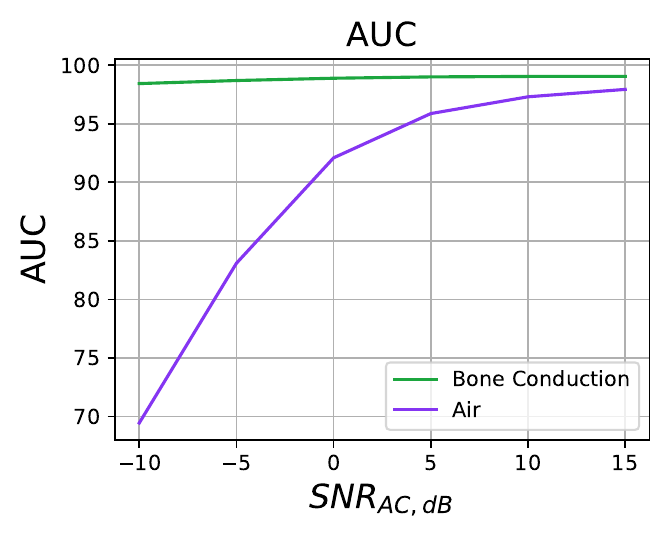}\\
        \vspace{-0.4cm}
        \subcaption{}
        \label{fig:mapped_snr_auc}
    \end{minipage}%
    \begin{minipage}[b]{0.3\textwidth}
	    \includegraphics[width=\textwidth]{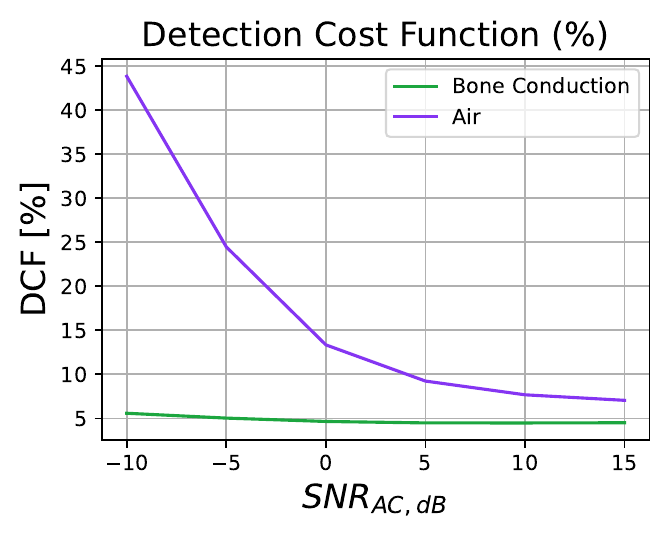}\\
        \vspace{-0.4cm}
        \subcaption{}
        \label{fig:mapped_snr_dcf}
    \end{minipage}%
    \begin{minipage}[b]{0.3\textwidth}
        \includegraphics[width=\textwidth]{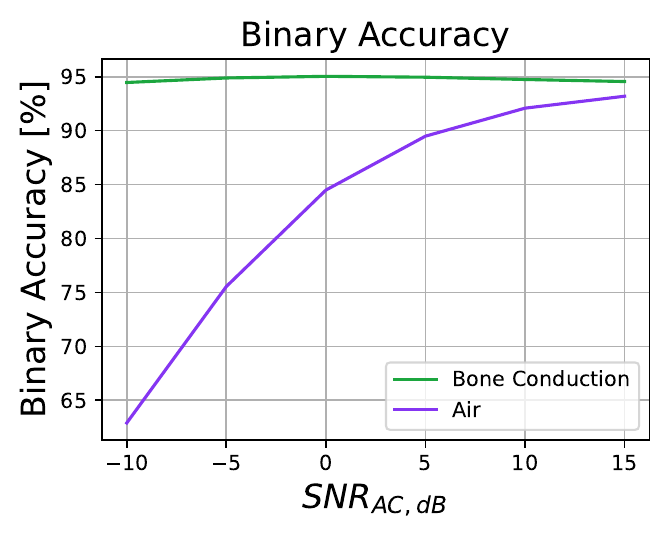}\\
        \vspace{-0.4cm}
        \subcaption{}
        \label{fig:mapped_snr_acc}
    \end{minipage}%
    \caption{Performance of BC-pVAD and AC-pVAD in terms of 
               AUC (\ref{fig:mapped_snr_auc}), 
               DCF (\ref{fig:mapped_snr_dcf}), and 
               Binary Accuracy (\ref{fig:mapped_snr_acc}) at a set 
               \boldmath
               $SNR_{AC, dB}$.}
               \unboldmath
    \label{fig:mapped_snr}
\end{figure*}

Using this method, both models were tested on a total of 5 hours of test data. The results of this 
evaluation are shown in table \ref{table:mapped_snr} and figure \ref{fig:mapped_snr}.

\begin{table}[b]
	\begin{tabular}{|l|l|l|l|}
	\hline
	$\boldsymbol{SNR_{AC, dB}}$ & \textbf{AUC}         & \textbf{DCF(\%)}   & \textbf{ACC}         \\
	\hline
	-10                          & 0.69 / \textbf{0.98} & 43 / \textbf{5.6}  & 0.63 / \textbf{0.94} \\
	\hline
	-5                           & 0.83 / \textbf{0.98} & 24 / \textbf{4.6}  & 0.75 / \textbf{0.94} \\
	\hline
	0                            & 0.92 / \textbf{0.98} & 13 / \textbf{4.5}  & 0.84 / \textbf{0.95} \\
	\hline
	5                            & 0.96 / \textbf{0.99} & 9.2 / \textbf{4.5} & 0.89 / \textbf{0.95} \\
	\hline
	10                           & 0.97 / \textbf{0.99} & 7.6 / \textbf{4.5} & 0.92 / \textbf{0.95} \\
	\hline
	15                           & 0.98 / \textbf{0.99} & 7.0 / \textbf{4.5} & 0.93 / \textbf{0.95} \\
	\hline
\end{tabular}

	\vspace{0.5cm}
    \caption{Performance of AC-pVAD / BC-pVAD in terms of AUC, DCF, and Binary Accuracy at a set 
           \boldmath$SNR_{AC, dB}$\unboldmath.}
	\label{table:mapped_snr}
\end{table}

\begin{figure*}[t]
    \begin{minipage}[b]{0.3\textwidth}
        \includegraphics[width=\textwidth]{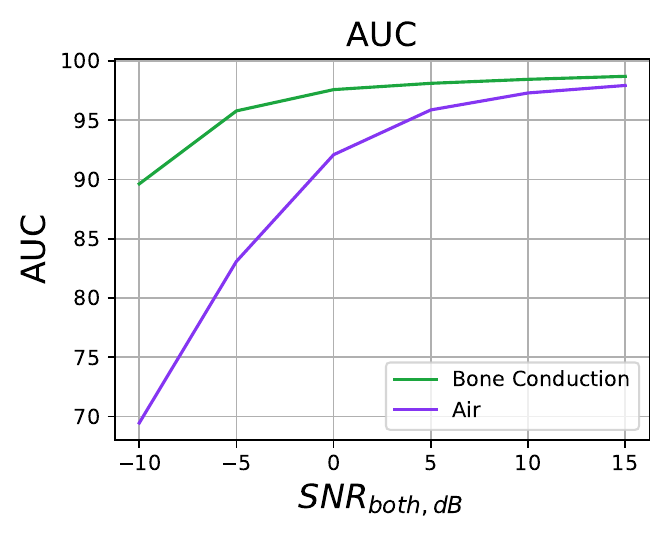}\\
        \vspace{-0.4cm}
        \subcaption{}
        \label{fig:same_snr_auc}
    \end{minipage}%
    \begin{minipage}[b]{0.3\textwidth}
	    \includegraphics[width=\textwidth]{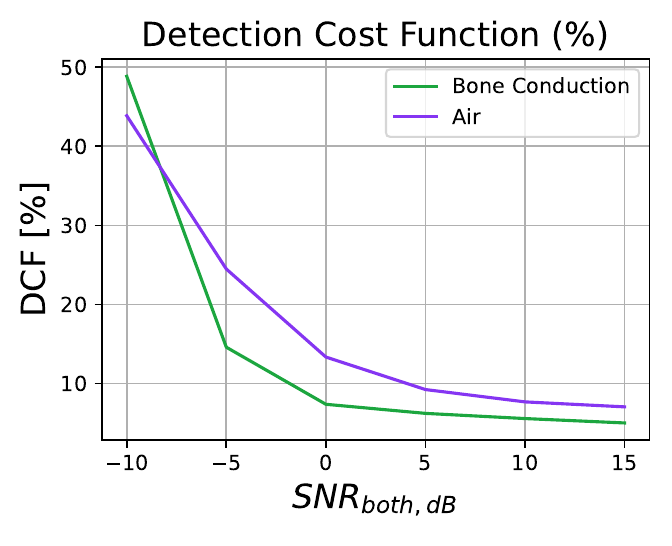}\\
        \vspace{-0.4cm}
        \subcaption{}
        \label{fig:same_snr_dcf}
    \end{minipage}%
    \begin{minipage}[b]{0.3\textwidth}
        \includegraphics[width=\textwidth]{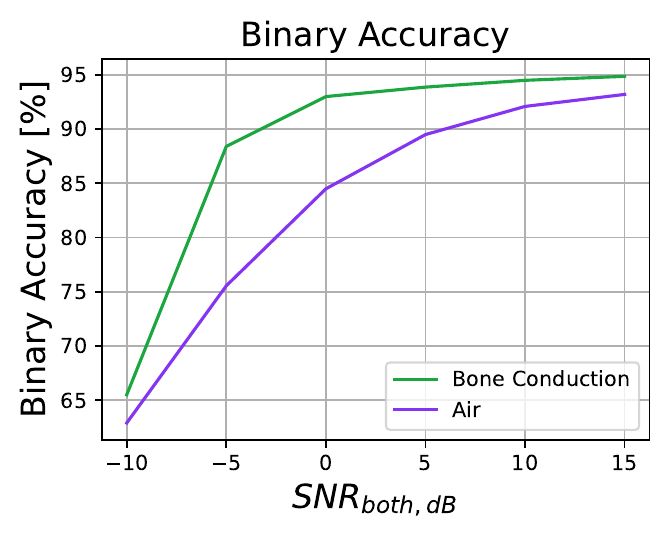}\\
        \vspace{-0.4cm}
        \subcaption{}
        \label{fig:same_snr_acc}
    \end{minipage}%
    \caption{Performance of BC-pVAD and AC-pVAD in terms of 
               AUC (\ref{fig:same_snr_auc}), 
               DCF (\ref{fig:same_snr_dcf}), and 
               Binary Accuracy (\ref{fig:same_snr_acc}) at a given
               \boldmath
               $SNR_{both, dB}$.}
               \unboldmath
    \label{fig:same_snr}
\end{figure*}

BC-pVAD achieves a binary accuracy of 94\% or better, at 0.98 AUC with a detection cost 
function of 5.6\%, with all metrics effectively independent of the 
surrounding noise. BC-pVAD continues to be highly effective at levels of interference 
where AC-pVAD ceases to be operational.

\paragraph{SNR Differences}

Table \ref{table:snr_delta} shows the difference in SNR between the air conduction and bone conduction 
data generated for the fixed $SNR_{AC, dB}$ evaluation of section \ref{sec:vad_eval}. On average, the bone-conduction 
microphone is capable of capturing the wearer's speech with an additional 15.01dB of SNR. 

\begin{table}[b]
	\begin{tabular}{|l|l|l|}
	\hline
	$\boldsymbol{SNR_{AC, dB}}$ & $\boldsymbol{SNR_{BC, dB}}$ & $\boldsymbol{SNR_{\Delta, dB}}$ \\
	\hline
    -10                         &    5.17                     &   15.17                          \\
	\hline
     -5                         &    10.09                    &   15.09                          \\
	\hline
      0                         &    15.01                    &   15.01                          \\
	\hline
      5                         &    19.92                    &   14.92                          \\
	\hline
     10                         &    24.84                    &   14.84                          \\
	\hline
\end{tabular}

	\vspace{0.5cm}
    \caption{Bone conduction SNR at different air conduction SNRs.}
	\label{table:snr_delta}
\end{table}


\paragraph{Evaluation at Equal SNR:}

To quantify how much of BC-pVAD's advantage is solely based on the higher SNR recording capabilities of 
the bone conduction microphone, the same analysis was repeated with both BC-pVAD and AC-pVAD 
being fed signals of equal SNR:

\begin{gather}
	y_{AC}(t) = \alpha_{AC}\cdot s_{AC}(t) + \beta_{AC}\cdot\eta_{AC}(t)\\%
	y_{BC}(t) = \alpha_{BC}\cdot s_{BC}(t) + \beta_{BC}\cdot\eta_{BC}(t)
\end{gather}
\begin{align}
	SNR_{both, dB} & = 10 \log_{10}\left(\frac{||\alpha_{AC}\cdot s_{AC}||^{2}}{||\beta_{AC}\cdot\eta_{AC}||^{2}}\right) \\ %
	               & = 10 \log_{10}\left(\frac{||\alpha_{BC}\cdot s_{BC}||^{2}}{||\beta_{BC}\cdot\eta_{BC}||^{2}}\right)
\end{align}

\begin{table}[b]
	\begin{tabular}{|l|l|l|l|}
	\hline
	$\boldsymbol{SNR_{both, dB}}$ & \textbf{AUC}         & \textbf{DCF(\%)}     & \textbf{ACC}         \\
	\hline
	-10                           & 0.69 / \textbf{0.90} & 43 /  \textbf{49}    & 0.63 / \textbf{0.66} \\
	\hline
	-5                            & 0.83 / \textbf{0.96} & 24 /  \textbf{14}    & 0.75 / \textbf{0.88} \\
	\hline
	0                             & 0.92 / \textbf{0.98} & 13 /  \textbf{7.4}   & 0.84 / \textbf{0.93} \\
	\hline
	5                             & 0.96 / \textbf{0.98} & 9.2 / \textbf{6.2}   & 0.89 / \textbf{0.94} \\
	\hline
	10                            & 0.97 / \textbf{0.98} & 7.6 / \textbf{5.6}   & 0.92 / \textbf{0.94} \\
	\hline
	15                            & 0.98 / \textbf{0.99} & 7.0 / \textbf{5.0}   & 0.93 / \textbf{0.95} \\
	\hline
\end{tabular}

	\vspace{0.5cm}
    \caption{Performance of AC-pVAD / BC-pVAD in terms of AUC, DCF, and Binary Accuracy  at a given
               \boldmath
               $SNR_{both, dB}$.}
               \unboldmath
	\label{table:same_snr}
\end{table}

These results are shown in table \ref{table:same_snr} and figure \ref{fig:same_snr}.
With a significant amount of additional noise to cope with, BC-pVAD's performance has been, as one would expect,
significantly reduced. 
While most metrics maintain comparable above $0dB$, performance below this point is reduced. At an SNR of -10dB, BC-pVAD is, capable 
of 0.9AUC, sees up to 49\% DCF, and achieves 66\% binary accuracy. In light of previous results showing a typical SNR difference of up to 15dB between the 
two microphones types, it should be noted that an $-10$dB  $SNR_{BC, dB}$ roughly corresponds to a $-25$dB $SNR_{AC, dB}$,
an impressively noisy environment.

Still, it outperforms AC-pVAD in almost every instance and metric, demonstrating that the network is still able to discern 
differences in the direct bone-conducted coupling of target speech and indirect coupling of external noise to the 
bone conduction microphone. 

%
\subsection{Quantization and Porting}

After the offline validation described in section \ref{sec:vad_eval}, the pVAD model 
was ported to run in real time on both the custom NRF5340 and custom Apollo 4 Blue platform described 
in section \ref{sec:hw}.

Both SoCs can interface with the Sonion VPU sensor using a standard PDM interface, which is converted 
to 16bit samples at 16kHz using hardware based PDM-to-PCM filters.

The pre-processing, as described in section \ref{sec:data_preproc}, was implemented 
using CMSIS-DSP, a highly optimized ARM-specific library, capable of exploiting features such as 
the core's SIMD (Single Instruction Multiple Data)
instructions to parallelize computation heavy operations such as the FFT. This has been shown to be up three times faster than a 
general purpose software FFT \cite{cmsis_dsp_fft}.

The BC-pVAD model developed in section \ref{sec:model_shape} and \ref{sec:model_train}, while small,
is still too computationally intensive to be evaluated in real time on a small embedded platform, due to 
its usage of floating point precision for all inputs, internal states, and outputs. 
Using Tensorflow, the model was quantized to 8bit precision, with a representative data set used to 
set internal gains and limit quantization error. Post quantization, the network lost at most 4\% of its original binary accuracy.

The inference is performed using TFlite Micro, a Tensorflow version targeted at edge computing, on both SoCs. 
TFlite is also able to utilize CMSIS-NN, another ARM-provided library for hardware accelerated calculations which 
has been shown to provide up to 4.6x improvement in throughput and 4.9x improvements in energy efficiency\cite{cmsis_nn}.

%
\subsection{Real-time Execution Analysis}

The processing time for the whole pVAD platform, averaged over 1024 executions
is detailed in table \ref{table:pwr}.

\begin{figure*}[htbp]
    \begin{minipage}[b]{0.45\textwidth}
    	\includegraphics[width=\textwidth]{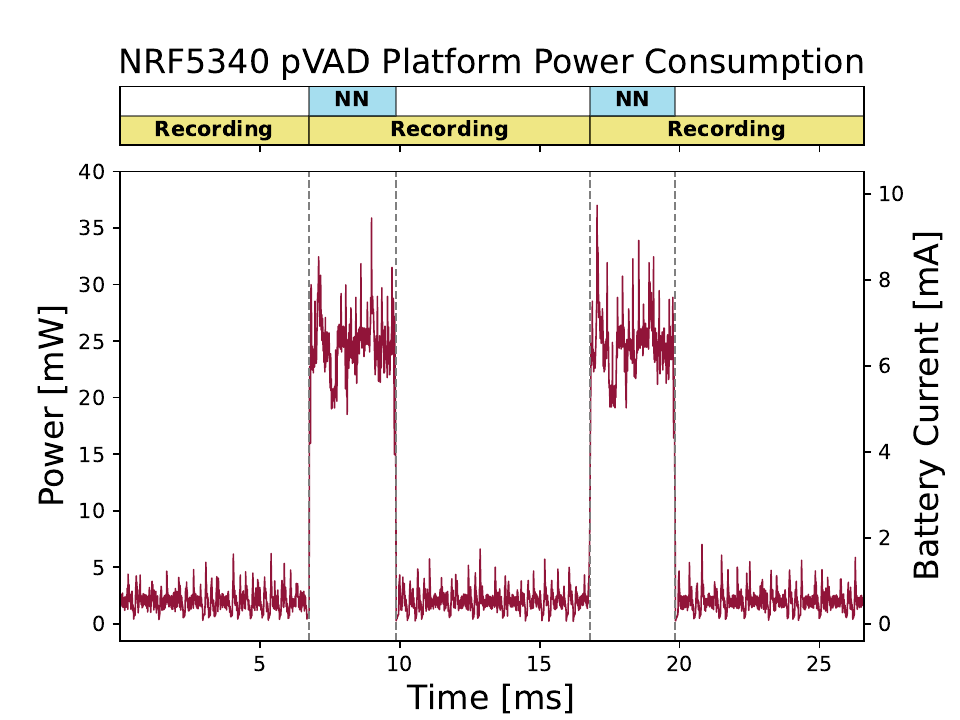}
        \vspace{-0.4cm}
        \caption{Power consumption of the NRF platform.}
    	\label{fig:pprofile_nrf}
    \end{minipage}%
    \hspace{0.05\textwidth}
    \begin{minipage}[b]{0.45\textwidth}
        \includegraphics[width=\textwidth]{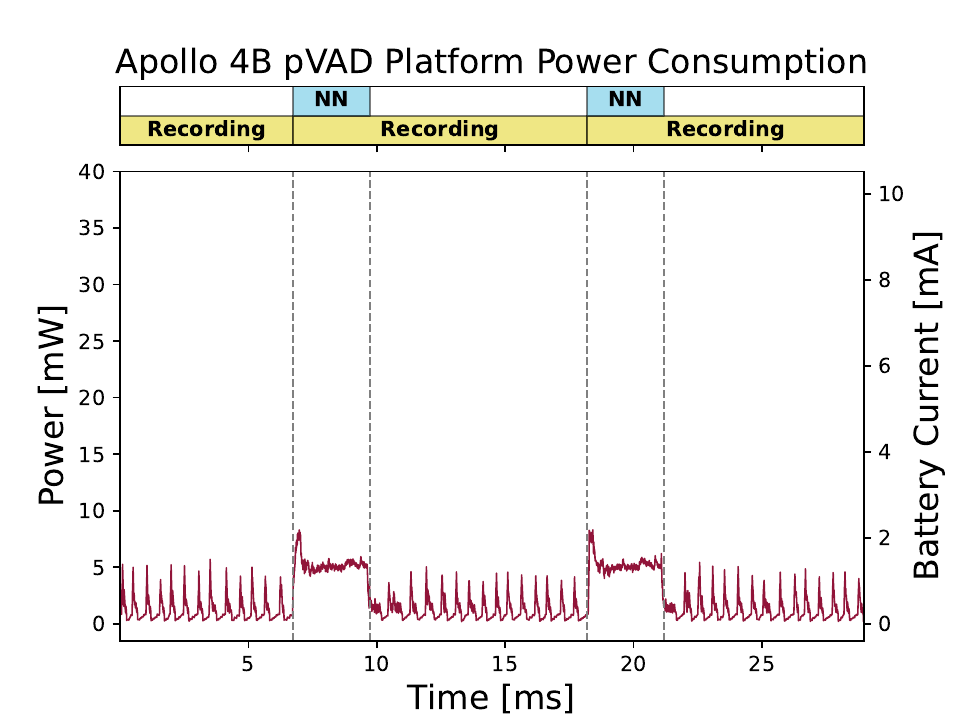}
        \vspace{-0.4cm}
        \caption{Power consumption of the Apollo platform.}
        \label{fig:pprofile_apo}
    \end{minipage}%
\end{figure*}

\begin{table}[b]
	\begin{tabular}{|l|l|r|r|}
	\hline
	\multicolumn{2}{|c|}{Processing Step}                      & NRF5340   & Apollo 4 Blue  \\
    \hline
    \multirow{2}{*}{Setup}             & memset                &  224.40us &    24.66us     \\
    \cline{2-4}
                                       & memcpy                &   60.22us &    12.81us     \\
	\hline
    \multirow{3}{*}{Pre-Processing}    & Hamming Window        &   29.92us &    28.41us     \\
    \cline{2-4}
                                       & RFFT                  &  249.03us &    419.34us    \\
    \cline{2-4}
                                       & $||\cdot||$           &  100.40us &    100.60us    \\
	\hline
    \multirow{2}{*}{Inference}         & memcpy                &   16.48us &     31.72us    \\
    \cline{2-4}
                                       & Inference             & 2313.90us &   2134.21us    \\
	\hline
	\multicolumn{2}{|r|}{\textbf{Total:}}                      & 2994.35us &   2803.40us    \\
    \hline
\end{tabular}

	\vspace{0.5cm}
    \caption{Processing time analysis and breakdown, averaged over 1024 inferences.}
	\label{table:pwr}
\end{table}

\paragraph{NRF5340:}
Running at 128MHz, the NRF5340 takes 2.99ms to process a single 320 sample frame. It 
spends 70.1\% of it's time in a low power sleep mode, and 29.9\% processing, with 
DMA-based data acquisition happening continuously at all times. Averaged over 1024 frames, 
2.065mW are consumed during sleep while acquiring microphone 
data. Of this, 1.97mW are consumed by the NRF, while 0.10mW are consumed by the microphone.
During computation the NRF's power consumption increases to 24.64mW, for a total consumption 
of 24.74mW. This power profile is shown in figure \ref{fig:pprofile_nrf}. On average the NRF 
platform consumes 9.20mW. With the installed 32mAh battery, this enables a theoretical battery life of 12.04h.

It should be noted that the NRF PDM-to-PCM converter's power consumption is very significant, making up a large
proportion of the power consumed during sleep.

\paragraph{Apollo 4 Blue:}
The Apollo, at 96MHz, requires 2.80ms to process a frame. Due to the 50\% overlap of 20ms frames, 
the platform has a maximum delay of 12.8ms between physical speech being recorded by the microphone 
and the next pVAD prediction becoming available.
Averaged over 1024 frames, 1.21mW are consumed during sleep, 1.01mW of which are consumed by the Apollo. 
During computation the Apollo requires an additional 5.01mW, for a total consumption 
of 5.11mW. This power profile is again shown in figure \ref{fig:pprofile_apo}. On average the Apollo 
platform consumes 2.64mW. On a 32mAh battery this enables a theoretical battery life of 43.10h.

This average battery current of 694 $\mu A$ at 3.8V corresponds to, assuming 95\% conversion efficiency, an average of 
1.393mA of SoC supply current at 1.8V.

Notably, this is significantly more than the 5$\mu A$/MHz or 0.48mA at 96MHz specified as
best-case operating current by the manufacturer for the Apollo 4 Blue. 
While this optimal operating current was achievable while running a manufacturer-provided benchmark firmware, 
it is not realistic to operate at such power levels with a more general purpose firmware.
The relatively large size of the model requires the usage of the far less power efficient SRAM region instead of
solely relying on the smaller TCM (tightly-coupled memory) as the benchmark does. This constant access to a
substantially larger memory region also negatively impacts the cache hit-rate, for which the optimal
benchmark was tuned. Nevertheless, the performance of the Apollo SoC is far superior to the NRF5340.

\begin{figure}[hbp]
	\centering
	\includegraphics[width=0.48\textwidth]{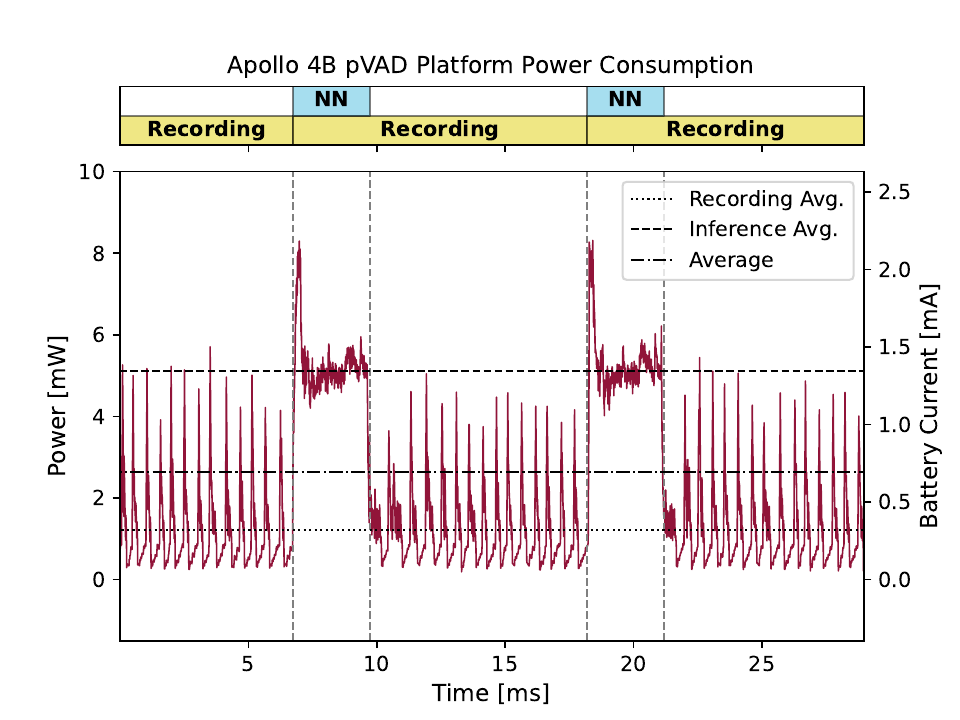}
	\caption{Power consumption profile of the Apollo platform.}
	\vspace{-2em}
\end{figure}


\subsection{Frame Qualification}

With the inference responsible for more than half of the total power consumed, 
much is to be gained by not only optimizing the power per inference, but also the 
number of inferences required.

While an inference is the most precise way of analyzing a given frame,
a heuristic rejecting frames with a very high probability of not containing 
any target speech has the potential of saving large amounts of power while 
having only a very negligible impact on pVAD performance. 
An implementation of such a frame pre-qualification scheme could be as simplistic as
basic thresholding.

\begin{figure*}[htbp]
    \begin{minipage}[b]{0.45\textwidth}
        \includegraphics[width=\textwidth]{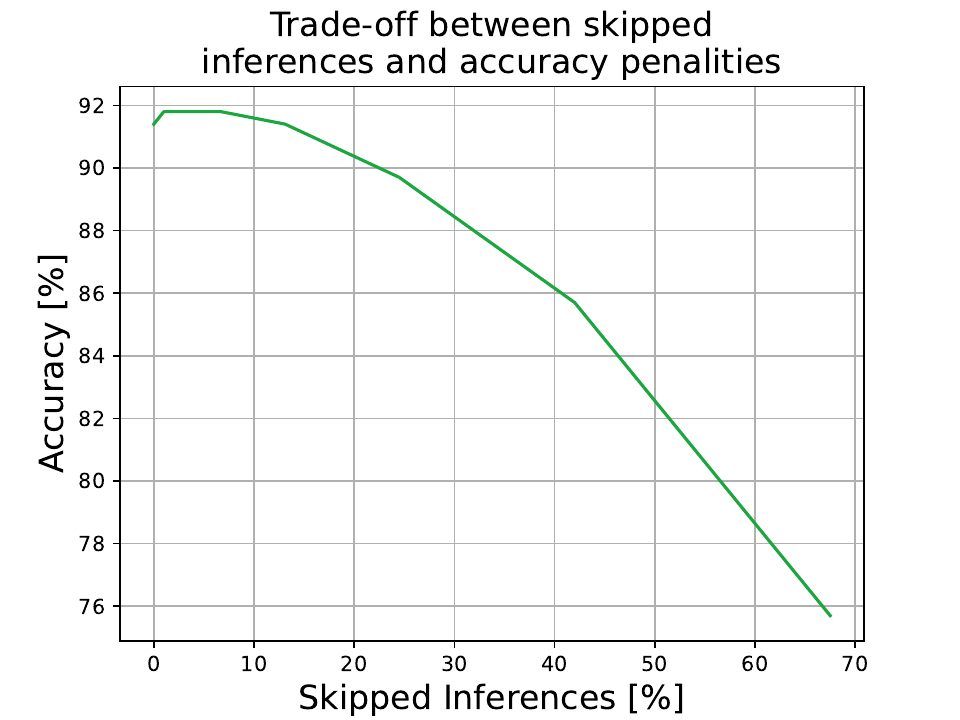}
        \vspace{-0.4cm}
        \caption{Relationship between percentage of inferences saved versus detection accuracy.}
        \label{fig:thresholding}
    \end{minipage}%
    \hspace{0.05\textwidth}
    \begin{minipage}[b]{0.45\textwidth}
        \centering
        \includegraphics[width=\textwidth]{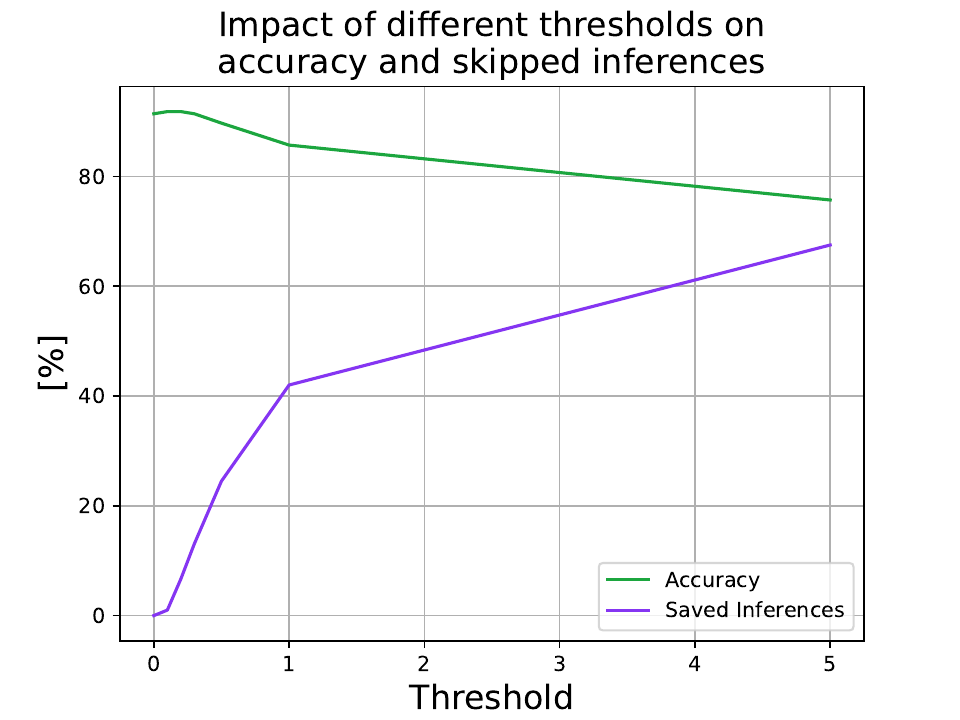}
        \vspace{-0.4cm}
        \caption{Changes in accuracy and percentage of inferences skipped at different basic thresholds.}
        \label{fig:thresholding_alt}
    \end{minipage}%
\end{figure*}

Due to the recurrent structure of the network used, such methodologies have to be 
carefully validated: Since the network bases its prediction not only on the frame 
containing target speech, but also it's potentially very long perceptive field 
of previous frames, skipping frames may have an adverse impact on the model's performance.
The (quantized) network's performance with a basic thresholding scheme in place was evaluated
on an input dataset with an SNR of 10 dB, with each clip normalized to a maximum amplitude of -15 dBFS.
With frames ignored if their' energy fell below a set threshold, the number of frames rejected and the 
overall accuracy of the network was recorded. This data is shown in figure \ref{fig:thresholding}.

\begin{figure}[htbp]
    \includegraphics[scale=0.50]{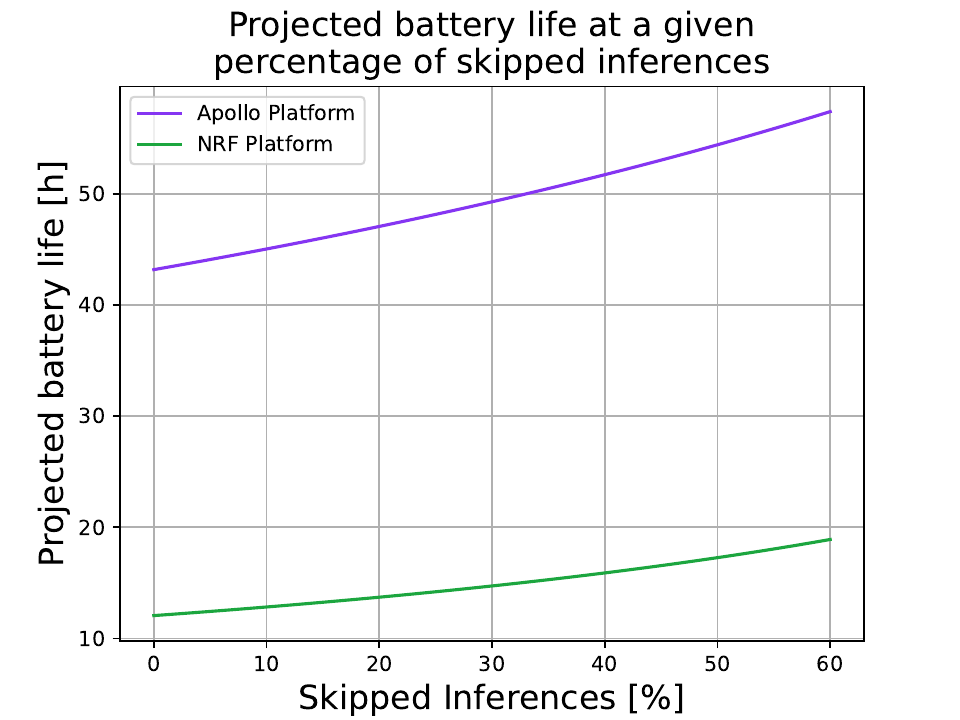}
    \centering
    \caption{Projection of total battery life on the NRF and Apollo platform at different percentages 
    of inferences skipped.}
    \label{fig:batlife}
\end{figure}

While high thresholds expectedly resulted in degraded performance as frames with important 
information were lost, a low threshold was capable of rejecting a significant number of 
frames with little to no impact on pVAD performance. A conservative threshold, at best,
is capable of preserving a large amount power in an environment with little noise, while, at 
worst, incurring only a small performance penalty at no extra cost power cost.

This motivates an evaluation of the estimated battery life of the device given a certain 
number of inferences that have been skipped based on some heuristic. While the actual number of 
skipped frames will certainly vary based both on the heuristic used and environment the user finds  
themselves in, it is not unrealistic to expect the device to experience intervals of silence 
in which many frames can be ignored.

This analysis, assuming that a frame can either be accepted or rejected after the FFT calculations,
is shown in figure \ref{fig:batlife} for both platforms. At 20\% of frames skipped, the Apollo platform
gains another 4h of battery life, while 40\% of frames skipped lead to 8h of additional battery
life.

	\section{Conclusions}

Bone conduction microphones enable many potential avenues in improving the quality of speech recorded using small true wireless headsets.
This paper quantifies some of the advantages that such microphones provide, 
demonstrating around 15dB of additional SNR over air conduction microphones 
mounted in the same earbud.

Furthermore, a personalized voice activity detection algorithm is present,
able to achieve 95\% binary accuracy consistently - even in exceedingly noisy environments leaving the air conduction microphone with only -10dB of SNR and 
unable to perform. The size of only approximately 5000 parameters makes this one 
of the smallest personalized voice activity detection networks available, especially compared to other models that do not require   
enrollment speech. This enables the developed algorithm to be run directly 
within the highly constrained environment of true wireless earbuds, a first, removing the 
need for slow and power-intensive data transfer. In this way, the complete system achieves an unprecedented high energy 
efficiency and fast response time for personalized VAD, at only 2.64mW consumed and 12.8ms between 
a speech event being recorded and predicted.

Motivated by the significant performance improvements seen by the use of bone conduction 
microphones, and the availability of energy-efficient, integrated, and flexible in-ear sensing
research platform developed in this paper, many more exciting bone conduction microphone applications can be
explored in the future:

The personalized voice activity detection demonstrated may enable the energy-efficient removal of not only 
random environmental noise, but also distractor speech that previously could not have been identified as 
interference.

While their tendency to impart an unnatural frequency response to the speech of the wearer makes them 
unattractive as a direct source of voice recording, their isolated perspective provides 
incredibly potent auxiliary data points. 
A combination of recent work in audio de-noising, such as RNNoise \cite{RNNoise} and PercepNet \cite{percepdsp}, 
with the bone conduction methods presented in this paper may enable a new generation of highly effective noise removal with minimal power requirements.

	\bibliographystyle{ACM-Reference-Format}
	\bibliography{bibliography}

\end{document}